\documentclass[aps,prb,twocolumn,showpacs,preprintnumbers,amsmath,amssymb,superscriptaddress]{revtex4}
\usepackage{mathptm}  
\usepackage{dcolumn}                    
\usepackage{bm}                        
\usepackage{graphicx}
\usepackage{times}
\usepackage{epstopdf}
\begin{document}

\newcommand{\Imag}{{\Im\mathrm{m}}}   
\newcommand{\Real}{{\mathrm{Re}}}   
\newcommand{\im}{\mathrm{i}}        
\newcommand{\talpha}{\tilde{\alpha}}
\newcommand{\ve}[1]{{\mathbf{#1}}}

\newcommand{\x}{\lambda}  
\newcommand{\y}{\rho}     
\newcommand{\T}{\mathrm{T}}   
\newcommand{\Pv}{\mathcal{P}} 
\newcommand{\vk}{\ve{k}} 
\newcommand{\vp}{\ve{p}} 

\newcommand{\N}{\underline{\mathcal{N}}} 
\newcommand{\Nt}{\underline{\tilde{\mathcal{N}}}} 
\newcommand{\g}{\underline{\gamma}} 
\newcommand{\gt}{\underline{\tilde{\gamma}}} 

\newcommand{\vecr}{\ve{r}} 
\newcommand{\vq}{\ve{q}} 
\newcommand{\ca}[2][]{c_{#2}^{\vphantom{\dagger}#1}} 
\newcommand{\cc}[2][]{c_{#2}^{{\dagger}#1}}          
\newcommand{\da}[2][]{d_{#2}^{\vphantom{\dagger}#1}} 
\newcommand{\dc}[2][]{d_{#2}^{{\dagger}#1}}          
\newcommand{\ga}[2][]{\gamma_{#2}^{\vphantom{\dagger}#1}} 
\newcommand{\gc}[2][]{\gamma_{#2}^{{\dagger}#1}}          
\newcommand{\ea}[2][]{\eta_{#2}^{\vphantom{\dagger}#1}} 
\newcommand{\ec}[2][]{\eta_{#2}^{{\dagger}#1}}          
\newcommand{\su}{\uparrow}    
\newcommand{\sd}{\downarrow}  
\newcommand{\Tkp}[1]{T_{\vk\vp#1}}  
\newcommand{\muone}{\mu^{(1)}}      
\newcommand{\mutwo}{\mu^{(2)}}      
\newcommand{\epsk}{\varepsilon_\vk}
\newcommand{\epsp}{\varepsilon_\vp}
\newcommand{\e}[1]{\mathrm{e}^{#1}}
\newcommand{\dif}{\mathrm{d}} 
\newcommand{\diff}[2]{\frac{\dif #1}{\dif #2}}
\newcommand{\pdiff}[2]{\frac{\partial #1}{\partial #2}}
\newcommand{\mean}[1]{\langle#1\rangle}
\newcommand{\abs}[1]{|#1|}
\newcommand{\abss}[1]{|#1|^2}
\newcommand{\Sk}[1][\vk]{\ve{S}_{#1}}
\newcommand{\pauli}[1][\alpha\beta]{\boldsymbol{\sigma}_{#1}^{\vphantom{\dagger}}}

\newcommand{\eq}{Eq.}
\newcommand{\eqs}{Eqs.}
\newcommand{\cf}{\textit{cf. }}
\newcommand{\ie}{\textit{i.e. }}
\newcommand{\eg}{\textit{e.g. }}
\newcommand{\etal}{\emph{et al.}}
\def\i{\mathrm{i}}

\title{Triplet supercurrent due to spin-active zones in a Josephson junction}

\author{Jacob Linder}
\affiliation{Department of Physics, Norwegian University of
Science and Technology, N-7491 Trondheim, Norway}

\author{Asle Sudb{\o}}
\affiliation{Department of Physics, Norwegian University of
Science and Technology, N-7491 Trondheim, Norway}

\date{\today}

\begin{abstract}

Motivated by a recent experiment evidencing  triplet superconductivity in a ferromagnetic Josephson junction with a Cu$_2$MnAl-Heusler 
barrier, we construct a theoretical model accounting for this observation. The key ingredients in our model which generate 
the triplet supercurrent are \textit{spin-active zones}, characterised by an effective canted interface magnetic moment. Using 
a numerical solution of the quasiclassical equations of superconductivity with spin-active boundary conditions, we find qualitatively 
very good agreement with the experimentally observed supercurrent. Further experimental implications of the spin-active zones are 
discussed.

\end{abstract}
\maketitle

Since the pioneering studies 50 years ago \cite{abrikosov}, the interplay between ferromagnetic (F) and superconducting (S) order has 
been much investigated, particularly so in recent years \cite{bergeret_rmp_05, buzdin_rmp_05}. This can largely be ascribed  to 
important experimental developments which have allowed for microscopic studies of both artificially engineered \cite{ryazanov} and 
intrinsic coexistence \cite{saxena} of F and S order, in addition to theoretical advances. One of the most exciting prospects in 
hybrid F$\mid$S structures is the possibility of tailoring the desired properties of the system on a nanometer scale. To accomplish 
this, it is necessary to take seriously the influence of the interface properties. Depending on whether the interfaces have 
spin-dependent properties or not, exotic new features may come into play in F$\mid$S structures, including long-range Josephson 
effects \cite{keizer_nature_06, eschrig_nphys_08, khaire_prl_10} and unconventional types of superconducting pairing 
\cite{berezinskii, volkov_prl_03, eschrig_jltp_07}.

In a very recent experiment by Sprungmann \etal~\cite{sprungmann_arxiv_10}, the importance of such interface properties was underscored. A 
long-range triplet supercurrent was observed in a S$\mid$F$\mid$S Josephson junction with a Cu$_2$MnAl-Heusler barrier acting as the F 
region. Surprisingly, Sprungmann \etal~observed a conventional, exponentially decaying supercurrent in the as-prepared state of the Cu$_2$MnAl 
layer, whereas a long-range supercurrent virtually independent of the junction thickness (up to a critical value) was observed upon annealing 
the junction. Beyond a critical value of the junction-thickness, an unusual abrupt decay of the supercurrent was observed. In the annealed 
case, the Heusler layer acquires a ferromagnetic order in its core, but retains spin glass order near the 
interfaces in the thickness range just above the onset of ferromagnetism. Therefore, it was suggested in Ref. \cite{sprungmann_arxiv_10} 
that the coupling between ferromagnetic and spin glass order would lead to a canted magnetization texture near the interfaces, which could 
be the necessary mechanism responsible for the observed triplet supercurrent. Illuminating this matter would be important to understand 
further the role of spin-active zones in Josephson junctions and their possible manipulation, which in turn could lead to tunable 
long-range supercurrents.

In this Rapid Communication, we construct a theoretical model to explain the experimental finding in Ref. \cite{sprungmann_arxiv_10} by including the 
role of canted magnetic interface moments in the quasiclassical theory of superconductivity. Employing a numerical solution in the diffusive 
regime of transport, we obtain very good qualitative agreement with observed supercurrent in Ref. \cite{sprungmann_arxiv_10}. Moreover, we 
highlight additional implications of spin-active zones in a Josephson junction. In particular, we find that if these zones couple together 
in a parallell alignment, \textit{the total supercurrent may actually vanish} for certain misalignment angles of the canted magnetic 
moments. Our results demonstrate the significance of spin-active zones in ferromagnetic Josephson junctions, and may provide a guideline 
for interpretation of experimental data and future investigations of such heterostructures.

\begin{figure}[t!]
\centering
\resizebox{0.40\textwidth}{!}{
\includegraphics{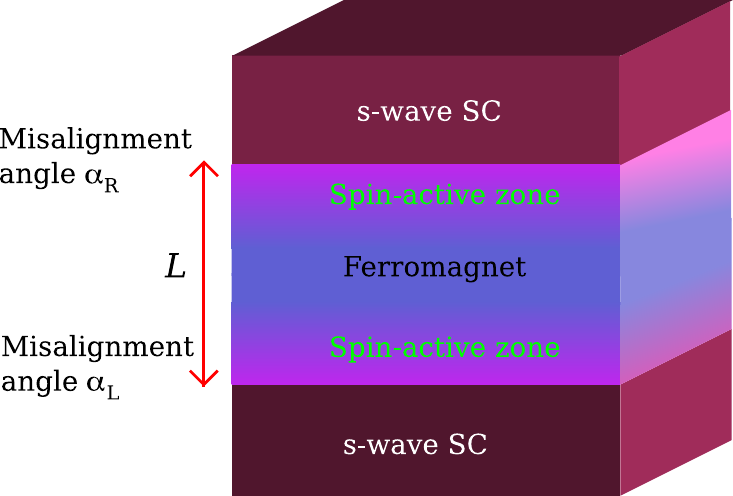}}
\caption{(Color online) The model employed for a Josephson junction with a ferromagnetic Heusler Cu$_2$MnAl barrier. The junction 
width is $L$, and we take into account a canted magnetization texture near the interfaces with misalignment angles $\alpha_{L,R}$ 
relative the  bulk magnetization. These spin-active zones generate a long-range supercurrent.}
\label{fig:model} 
\end{figure}

The system under consideration is shown in Fig. \ref{fig:model}. A ferromagnetic region of width $L$ is sandwiched between two standard 
$s$-wave superconductors (\eg Nb). Near the interfaces, we identify \textit{spin-active zones} which may arise \eg due to the presence 
of magnetic disorder or canted magnetic moments which are misaligned compared to the bulk magnetization direction. To study this system, 
we employ the quasiclassical theory of superconductivity which provides equations of motion for the Green's functions of the system. By 
supplementing these equations with proper boundary conditions for spin-active interfaces with possibly misaligned magnetic moments, we 
are able to compute the supercurrent flowing through the system. We give a brief account of the theoretical framework here, and refer 
the reader to \eg Ref. \cite{bergeret_rmp_05} for a comprehensive review. In the F region, we employ a Ricatti-parametrization 
\cite{schopohl_prb_95} of the Green's function as follows
\begin{align}
\hat{g} &= \begin{pmatrix}
\N(\underline{1} - \g\gt) & 2\N\g \\
2\Nt\gt & \Nt(-\underline{1} + \gt\g) \\
\end{pmatrix}, 
\end{align}
where $\N = (\underline{1}+\g\gt)^{-1}$ and $\Nt = (\underline{1}+\gt\g)^{-1}$. It also satisfies $(\hat{g})^2 = \hat{1}$, and is 
characterized by the two unknown $2\times2$ matrices $\g$ and $\gt$. The above Green's functions satisfy the Usadel 
equation \cite{usadel_prl_70} 
\begin{align}
\mathcal{D}\hat{g} \partial_x\hat{g}+ i[\varepsilon\hat{\rho}_3 + h\text{diag}(\underline{\tau_3},\underline{\tau_3}), \hat{g}] = 0.
\end{align}
Here, $\mathcal{D}$ is the diffusion coefficient in the F region, $h$ is the magnitude of the exchange field, while $\varepsilon$ is 
the quasiparticle energy. To obtain a complete solution of the Green's function, it is necessary to supplement the Usadel equation 
with boundary conditions \cite{spinactive}. Assuming that the F region extends from $x=0$ to $x=L$, the boundary conditions read
\begin{align}\label{eq:bc}
2\gamma_T^lL\hat{g}\partial_x\hat{g} = [\hat{g}^l,\hat{g}] + \i\gamma_s^l[\hat{M}(\alpha_L),\hat{g}] \text{ at } x = 0,\notag\\
2\gamma_T^rL\hat{g}\partial_x\hat{g} = [\hat{g}, \hat{g}^r] - \i\gamma_s^r[\hat{M}(\alpha_R),\hat{g}] \text{ at } x = L.
\end{align}
Here, we have introduced
\begin{align}
\hat{M}(\alpha) = (\cos\alpha)\text{diag}(\underline{\sigma_z}, \underline{\sigma_z}) + (\sin\alpha)\text{diag}(\underline{\sigma_y}, \underline{\sigma_y}^*) 
\end{align}
 with $\phi$ being the angle between the barrier magnetic moment and the bulk magnetization ($z$-axis) as indicated in Fig. \ref{fig:model}. 
 Above, $\underline{\sigma_j}$ denotes the j'th Pauli matrix in spin space. We have defined $\gamma_T^{l,r} = R_B^{l,r}/R_F$ and 
 $\gamma_s^{l,r} = 1/(R_NG_s^{l,r})$, where $R_N$ is the normal-state resistance of the N region, $R_B^{l,r}$ is the barrier resistance 
 at the left/right interface, while $G_s^{l,r}$ accounts for the spin-dependent interfacial phase-shifts at the left/right interface. 
 In the S region, we make use of the bulk solution for the Green's function
\begin{align}
\hat{g}^{l,r} &= \begin{pmatrix}
c & 0 & 0 & s\e{\pm\i\chi/2} \\
0 & c & -s\e{\pm\i\chi/2} & 0 \\
0 & s\e{\mp\i\chi/2} & -c & 0 \\
-s\e{\mp\i\chi/2} & 0 & 0 & -c \\
\end{pmatrix},
\end{align}
where $c=\cosh\theta$, $s=\sinh\theta$, $\theta=\text{atanh}(|\Delta_0|/\varepsilon)$, and the superconducting phase difference is $\chi$. 
This approximation is valid for low transparency interfaces in which case the proximity effect is weak. The above equations constitute a 
closed set which may be solved numerically, and we add a small imaginary part to the quasiparticle energy for improved numerical 
stability, i.e. $\varepsilon \to \varepsilon + \i\delta$ with $\delta/\Delta_0=0.01$. To model the experimental situation in 
Ref. \cite{sprungmann_arxiv_10}, we consider the weak proximity effect regime with a low-barrier transparency and an
exchange field $h$. Specifically, we choose $\gamma_T^{l,r}=40$ and $h/\Delta_0=80$ which is within the regime of validity for 
quasiclassical theory when assuming $\Delta_0\simeq 1$ meV. Also, in this regime one may safely neglect magnetoresistance terms $\gamma_\text{MR}$ in the boundary conditions \cite{spinactive}. We have chosen a numerical approach here to avoid overburdening the paper with cumbersome analytical expressions.

It was proposed by Sprungmann \etal~  \cite{sprungmann_arxiv_10} that the microscopic mechanism generating the triplet supercurrent 
was the specific magnetization profile in the Cu$_2$MnAl layer, featuring a canted magnetization texture near the interface regions. 
An important observation in Ref. \cite{sprungmann_arxiv_10} was that the triplet supercurrent was only observed in a limited thickness 
range, and died off rapidly 
above a critical thickness $L_c$ thus leaving behind only the conventional short-ranged Josephson current. This indicates that the 
misalignment angle of the interface magnetization compared to the bulk depends on the thickness $L$ of the layer. Such a 
conjecture agrees with the fact that the Heusler layers gradually make a transition near the interfaces from pure spin-glass 
to coexisting spin-glass order with a ferromagnetic moment of increasing magnitude as the thickness increases. To phenomenologically
model such behavior, we write the misalignment angles shown in Fig. \ref{fig:model} as
\begin{align}\label{eq:alpha}
\alpha = \alpha_0/[1 + \e{\zeta(L-L_c)}],
\end{align}
where $\alpha_0$ is the misalignment angle at the onset of ferromagnetic order and $\zeta\ll1$ accounts for the slope with which the 
canted moments relax into the same orientation as the bulk magnetization. 

Let us further qualify \cite{sprungmann} the model employed here for the canted interface magnetization texture. A thin film Heusler alloy features a spin glass structure if the film is sputtered at room temperature. Annealing will induce a transposition of nearest-neighbour atoms and tries 
to drive the compound into a ferromagnetically coupled structure. The maximum ordering effect is achieved within the core 
of the layer, whereas at the interfaces this ordering process is disturbed 
due to \eg interdiffusion. Here, an antiferromagnetic 
Mn-Mn-coupling will persist locally and competes with the ferromagnetic 
ordering. This causes the mentioned canting of the magnetization axis 
defined by the one of the core. The issue of what the 
magnetization profile looks like at the interfaces of thin Heusler layers was investigated experimentally in Ref. \cite{bergmann_prb_05}. We here also briefly mention a theoretical work on the influence of disorder in Co-based Heusler 
alloys by Picozzi \etal \cite{picozzi_prb_04}, although the comparison should be considered carefully since in Co-based 
Heuslers the Co contributes effectively to ferromagnetism whereas Cu does not carry any magnetic moment in the relevant structure.

\begin{figure}[t!]
\centering
\resizebox{0.45\textwidth}{!}{
\includegraphics{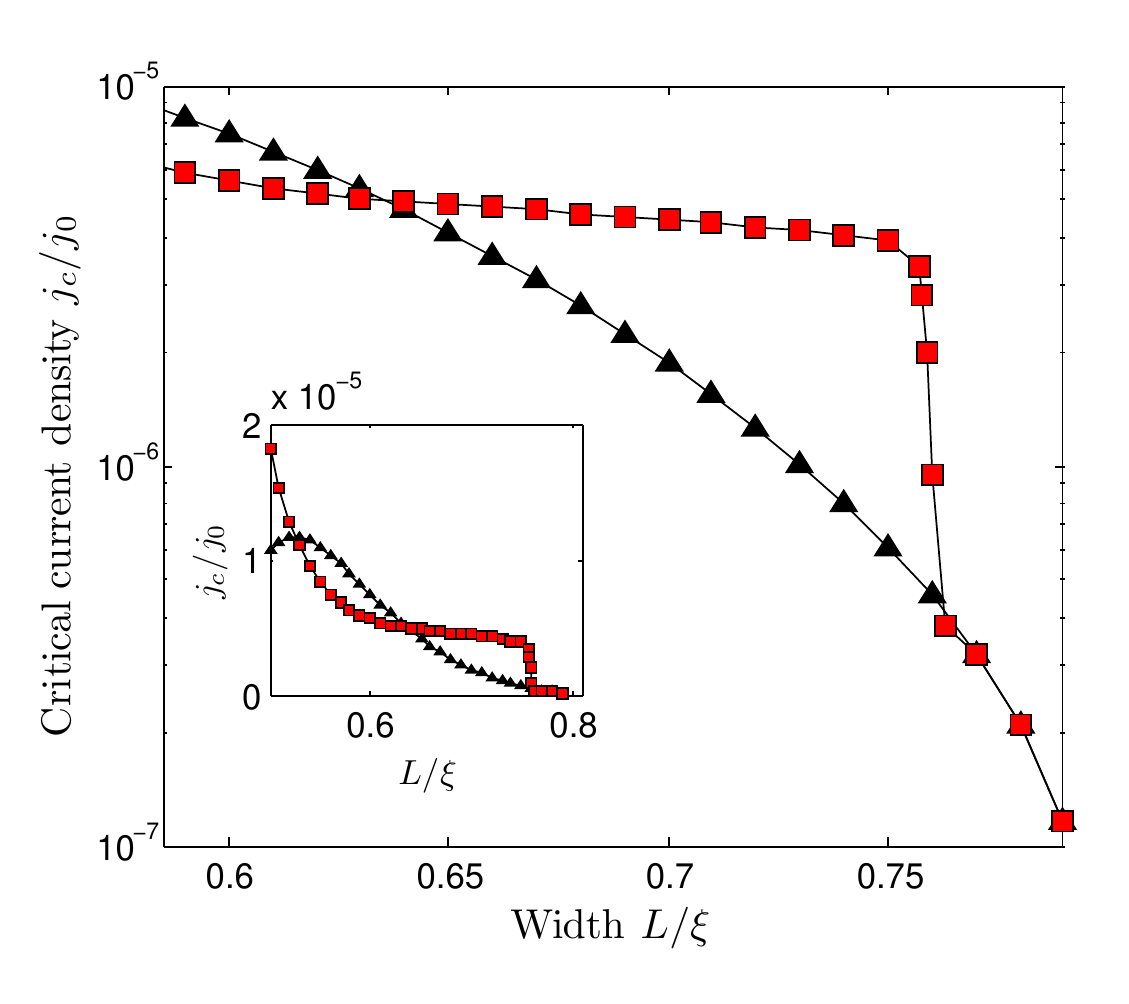}}
\caption{(Color online) Plot of the critical current density vs. the junction width $L$. Triangles (black) represent results 
for the case $\alpha=0$, i.e. no canted interface moments. Squares (red) represent results for misalignment angle $\alpha$ 
given by Eq. \ref{eq:alpha}. We have set $\alpha_L=\alpha_R$ with $\alpha_0=\pi/4$ and $\gamma_\phi^{l,r}=15$. Assuming a 
superconducting coherence length of $\xi\simeq 14$ nm in dirty Nb \cite{sangiorgio_prl_08}, we use $L_c/\xi=0.76$ to obtain 
the sharp drop-off near $L=10.5$ nm as in Ref. \cite{sprungmann_arxiv_10}. \textit{Inset:} Zoom-out version of the current 
density.}
\label{fig:main} 
\end{figure}

We are now in a position to calculate the normalized critical current density of the junction, which in the weak-proximity 
effect regime is obtained via
\begin{align}
j_c/j_0 &= \max_{\chi} \Big|\int^\infty_0 \text{d}\varepsilon \tanh\frac{\beta\varepsilon}{2} \sum_j \text{Re}\{\mathcal{M}_j\}\Big|,\; j=\{\pm,\sigma=\uparrow,\downarrow\},
\end{align}
where we have defined 
\begin{align}
\Delta_0M_\pm/\xi &= [f_\pm(-\varepsilon)]^*\partial_x f_\mp(\varepsilon) - f_\pm(\varepsilon)\partial_x[f_\mp(-\varepsilon)]^*,\notag\\
\Delta_0M_\sigma/\xi &= [f_\sigma(-\varepsilon)]^*\partial_x f_\sigma(\varepsilon) - f_\sigma(\varepsilon)\partial_x[f_\sigma(-\varepsilon)]^*.
\end{align} Here, $\{f_t,f_s,f_\uparrow,f_\downarrow\}$ denote the $S_z=0$ triplet, singlet, and equal spin-pairing anomalous Green's 
function induced in the F region, and we have defined $f_\pm = f_t \pm f_s$. To make direct contact with the experiment of Ref.
 \cite{sprungmann_arxiv_10}, we contrast a situation with $\alpha=0$, i.e. no canted interface moments, against a situation where the 
 misalignment angle $\alpha$ evolves with the thickness $L$ according to Eq. (\ref{eq:alpha}). The result is shown in Fig. \ref{fig:main}. 
 The result is seen to be qualitatively in very good agreement with the finding or Ref. \cite{sprungmann_arxiv_10}: the supercurrent shows 
 little decay upon increasing the thickness $L$ in the regime where the spin-active zones generate a misalignment angle, and then 
 collapses onto the conventional singlet result above a certain thickness $L_c$. This suggests that the spin-dependent interface 
 properties may play a pivotal role in the generation of the triplet supercurrent. In our picture, the abrupt change in the 
 spin-triplet Josephson current is due to an abrupt change in the angle $\alpha$ in Eq. \ref{eq:alpha} 
 
 \begin{figure}[t!]
\centering
\resizebox{0.45\textwidth}{!}{
\includegraphics{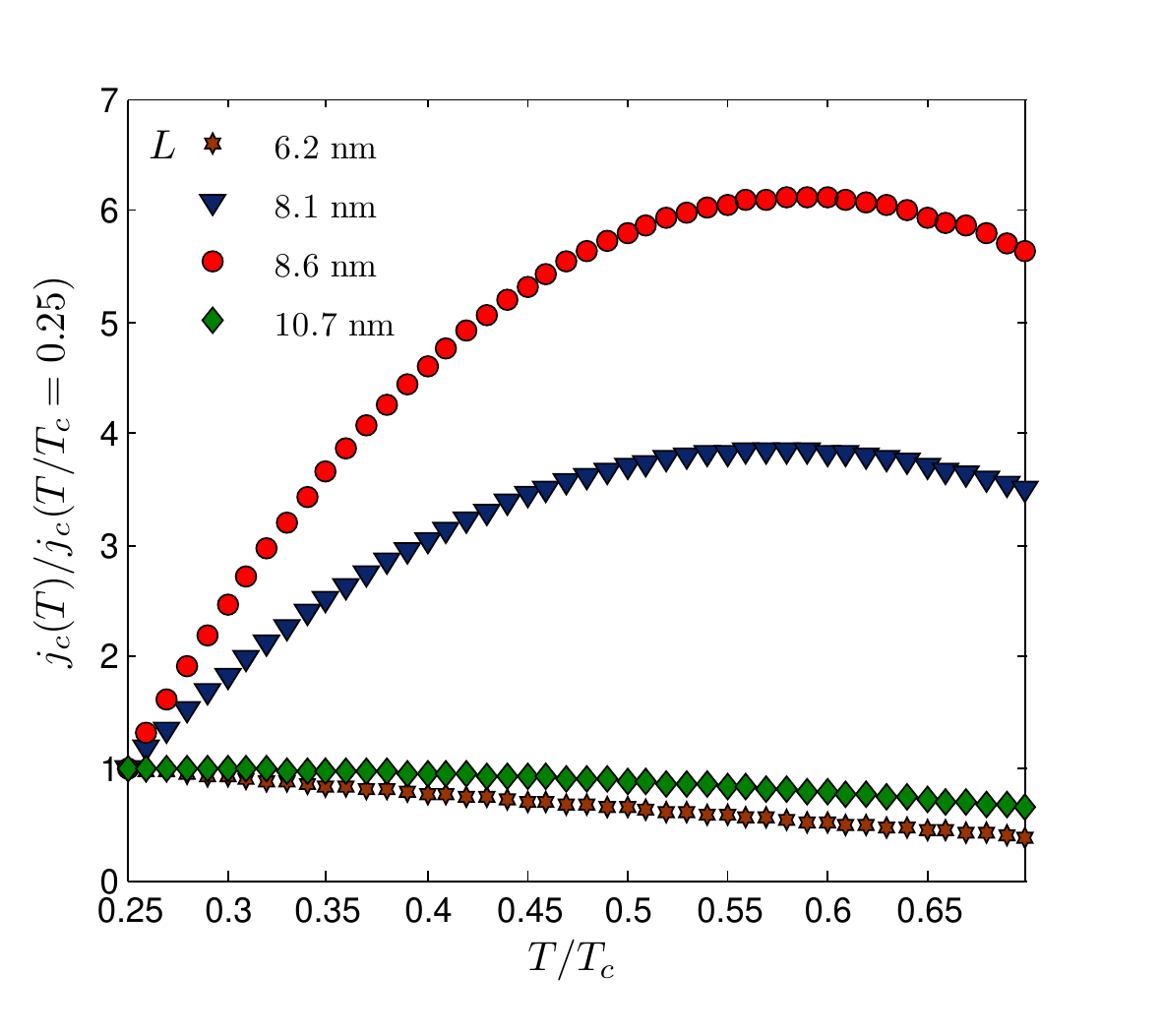}}
\caption{(Color online) Plot of the critical current density vs. temperature. To make contact with the experiment in Ref. \cite{sprungmann_arxiv_10} where $T_c\simeq 8$ K, we normalize the current to its value at $T/T_c=0.25$. The rest of the parameters are as in Fig. \ref{fig:main}.}
\label{fig:vsT} 
\end{figure}

Another interesting finding in Ref. \cite{sprungmann_arxiv_10} is the non-monotonic temperature dependence of the critical current observed in the region of widths $L$ where the long-range current is dominant. With $T_c\simeq 8$ K being the critical temperature \cite{sprungmann}, we plot the current vs. temperature in Fig. \ref{fig:vsT} for three choices of the width: (i) right below the region of a dominant triplet current ($L=6.1$ nm), (ii) in the middle of the region of a dominant triplet current ($L=8.1$ nm and $8.6$ nm), and (iii) right after the vanishing long-range current ($L=10.7$ nm). We again obtain a good match to the results of Ref. \cite{sprungmann_arxiv_10}. It should be mentioned that we do find signs of 0-$\pi$ oscillations for some particular choices of intermediate widths $L$, but the non-monotonic behavior in the regime of a dominant long-range current is nevertheless confirmed. Close examination of the current vs. temperature curves in Ref. \cite{sprungmann_arxiv_10} reveals a hint of small-scale oscillation superimposed on the overall non-monotonic behavior for some particular widths $L$. We were not able to reproduce these fine-scale oscillations within our model, and speculate that these might pertain to a more complicated magnetization profile in the bulk of the Cu$_2$MnAl layer since similar behavior has been predicted \cite{halasz_prb_09} in the conical ferromagnet Ho. We intend to address this issue in a forthcoming work. It should be noted that the spin-dependent phase-shifts occuring at the interfaces can in general influence the 0-$\pi$ transition pattern both as a function of the width and temperature of magnetic as well as non-magnetic (normal interlayer) Josephson junctions.  

\begin{figure}[t!]
\centering
\resizebox{0.45\textwidth}{!}{
\includegraphics{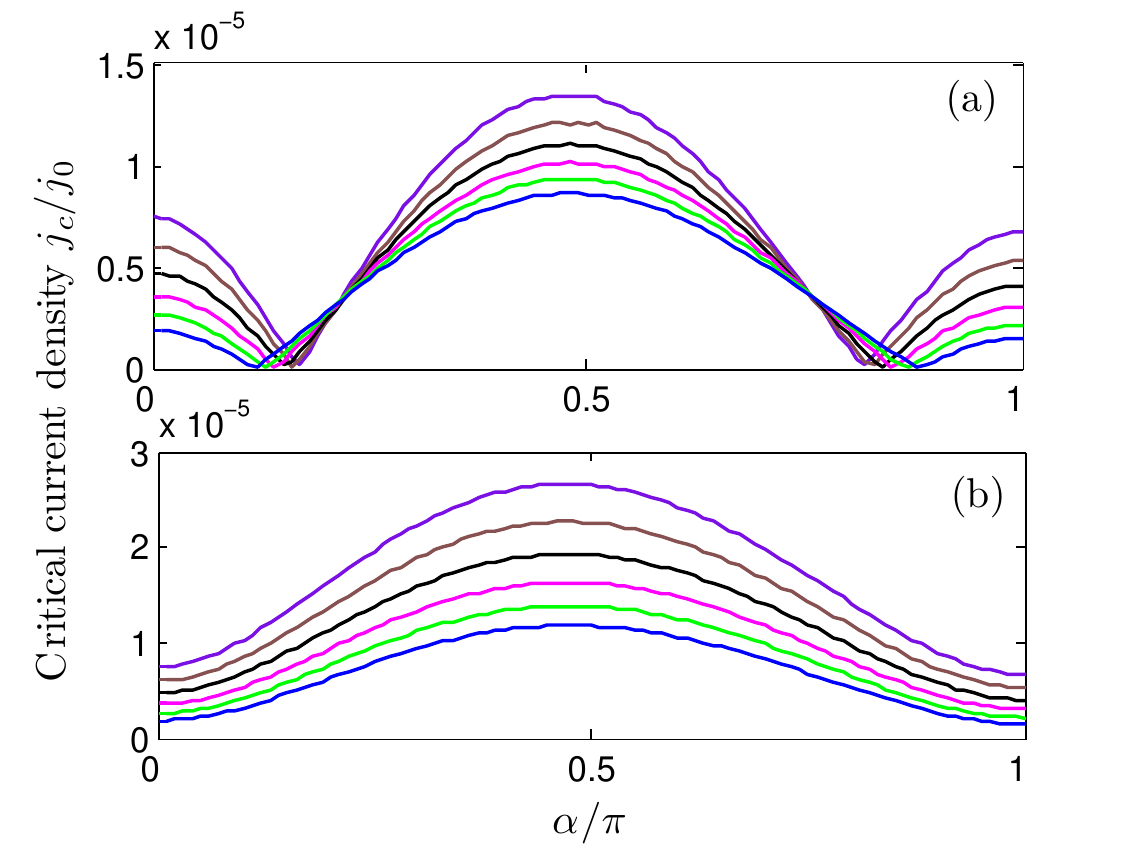}}
\caption{(Color online) Plot of the critical current density vs. the misalignment angle $\alpha$ in the spin-active zones. In (a), 
we set $\alpha_L = \alpha_R$ while in (b) $\alpha_L = -\alpha_R$. From top to bottom at $\alpha=\pi/2$, the curves correspond to 
thicknesses $d_F/\xi=\{0.60,0.62,0.64,0.66,0.68,0.70\}$. }
\label{fig:vsalpha} 
\end{figure}

To further highlight the influence on the current density by the presence of a canted magnetization texture near the interfaces, we plot 
in Fig. \ref{fig:vsalpha} the current density vs. the misalignment angle in the case of an (a) parallel coupling $\alpha_L = \alpha_R$ 
between the spin-active zones and an (b) antiparallel coupling $\alpha_L = -\alpha_R$. The main difference between these two cases is 
that the total supercurrent may actually \textit{vanish} at some misalignment angles when the coupling is parallel. In this scenario, 
the contribution from the long-range triplet current effectively cancels out the singlet supercurrent. Common for both scenarios is 
that the maximally attainable critical current occurs near $\alpha=\pi/2$. However, it should be noted that the maximum does not occur 
exactly at $\alpha=\pi/2$, but rather at a slight off-set $\alpha_c < \pi/2$. When we reverse the bulk magnetization direction, i.e. 
$h\to(-h)$, we note that the maximum of the critical current occurs at an angle $\alpha_c>\pi/2$ which is equidistant from $\pi/2$ 
compared to the former case. This behavior can be understood from a symmetry perspective. Namely, the system is not invariant under 
a spatial inversion whenever $h\neq0$ due to the definite direction of the magnetization. Therefore, the angle $\alpha_c$ providing 
the maximum critical current will either be smaller or larger than $\pi/2$, depending on whether $h$ points along $\hat{\mathbf{z}}$ 
or $-\hat{\mathbf{z}}$.

So far, we have constructed a theoretical model for the spin-active zones in the Heusler layer which is able to account well for the 
experimental finding in Ref. \cite{sprungmann_arxiv_10}. There are nevertheless extensions of this model which could be appropriate 
to pursue in order to further clarify the underlying physics, and which we comment on. In our treatment, we have 
included the spin-active zones as effective canted magnetization textures near the interface and included the resulting conductance-terms 
in the quasiclassical boundary conditions. This is certainly only an effective model for the interplay between spin-glass and 
ferromagnetic order in the Heusler layers in the interface region. In a  more microscopic lattice-model treatment of the system, one could 
address how the specific details of such an interplay would influence the critical current behavior. Another issue of
interest, is the precise role of the spatially dependent magnetization texture in the ferromagnetic layer. Recent 
theoretical works \cite{longrange} have highlighted the necessary requirements of the magnetization profile which renders possible 
a long-range triplet current. It would thus be interesting to experimentally extract this profile and its dependence on both spatial 
coordinate and its evolution in terms of magnetization direction, so that one could construct a more accurate model for this system. 
Qualitatively, we expect that our model captures the essential features which could account for the generation of a long-range 
triplet supercurrent in the experimental system of Ref. \cite{sprungmann_arxiv_10}.

In summary, we have constructed a theoretical model to account for recent experimental findings in Ref. \cite{sprungmann_arxiv_10} which 
presented evidence for triplet superconductivity in a ferromagnetic Josephson junction with a Cu$_2$MnAl-Heusler barrier. The crucial 
ingredients in our model which generate the triplet supercurrent are \textit{spin-active zones} near the interfaces due to a coupling 
between ferromagnetic and spin glass order, resulting in an effective canted magnetization texture near the interfaces. Using a 
numerical solution of the quasiclassical equations of superconductivity with spin-active boundary conditions, we find qualitatively 
very good agreement with the experimentally observed supercurrent. Further experimental implications of the spin-active zones have 
also been discussed.

\textit{Acknowledgments.} The authors acknowledge support by the Norwegian Research Council, Grant No. 167498/V30 (STORFORSK), and thank D. Sprungmann for a very useful exchange.

\end{document}